\documentclass[preprint, apl]{revtex4}

\usepackage{graphicx}
\usepackage{amsmath}
\usepackage{amssymb}
\usepackage{amsfonts}

\renewcommand{\deg}{$^\circ$}

\begin{document}
\title{On the extraction of resistivity and area of nanoscale interconnect lines by temperature-dependent resistance measurements}
\author{Christoph Adelmann}
\email{christoph.adelmann@imec.be}
\address{Imec, 3001 Leuven, Belgium}

\begin{abstract}
Several issues concerning the applicability of the temperature coefficient of the resistivity (TCR) method to scaled interconnect lines are discussed. The central approximation of the TCR method, \emph{i.e.} the substitution of the interconnect wire TCR by the bulk TCR becomes doubtful when the resistivity of the conductor metal is strongly increased by finite size effects. Semiclassical calculations for thin films show that the TCR deviates from bulk values when the surface roughness scattering contribution to the total resistivity becomes significant with respect to grain boundary scattering, an effect that might become even more important in nanowires due to their larger surface-to-volume ration. In addition, the TCR method is redeveloped to account for line width roughness. It is shown that for rough wires, the TCR method yields the harmonic average of the cross-sectional area as well as, to first order, the accurate value of the resistivity at the extracted area. Finally, the effect of a conductive barrier or liner layer on the TCR method is discussed. It is shown that the liner or barrier parallel conductance can only be neglected when it is lower than about 5 to 10\%{} of the total conductance. It is furthermore shown that neglecting the liner/barrier parallel conductance leads mainly to an overestimation of the cross-sectional area of the center conductor whereas its resistivity is less affected.\bigskip

\end{abstract}

\maketitle

\section{Introduction}

Interconnects are fundamental elements of any microelectronic circuit since they distribute signals as well as provide power and grounding for various components of the circuit. The continuing miniaturization of microelectronic circuits following Moore's law relies crucially on the scaling of the interconnect pitch \cite{M_2003, CGJ_2014,BAZ_2015}. In recent technology nodes, interconnect wires have reached critical dimensions around 20\,nm and, in the near future, dimensions of 10\,nm and below are potentially envisaged. At such scales, the interconnect resistance per unit length, a key performance parameter, increases strongly not only due to the smaller cross-sectional area but also due to a degraded conductor conductivity \cite{JBT_2009}. This effect is due to augmented scattering of the charge carriers by surfaces \cite{F_1938,S_1952,S_1967} as well as grain boundaries \cite{MSJ_1969,MS_1970}, as grain sizes are typically reduced in narrower interconnect lines. Moreover, low-conductivity barrier and liner layers, required to ensure interconnect reliability, occupy an increasing wire volume fraction, increasing the overall line resistance further. Therefore, the interconnect performance has become a key constraint for the overall circuit performance, limiting further progress in circuit miniaturization \cite{CGJ_2014,I_2004,TCR_2016}. To mitigate this effect, alternatives to Cu as conductor materials have been researched with increasing intensity for several years \cite{TCR_2016,KMS_2002,AWP_2014,PN_2014}. Ideal alternative metals show a much weaker thickness dependence of the resistivity than Cu and allow for reliable operation without the need for barrier and liner layers. 

While the final performance of an interconnect is determined by the line \emph{resistance}, the understanding of the interconnect behavior often requires the measurements of the \emph{resistivity} (or the conductivity) of the conductor material. While the experimental determination of the resistivity of thin metallic films from sheet resistances is straightforward when the film thickness is known, the determination of the resistivity of narrow metallic wires is more complex as the accurate measurement of the cross-sectional wire area is often difficult. The physical area of interconnects can \emph{e.g.} be determined by transmission-electron microscopy (TEM). However, the low throughput of TEM does not allow for a statistical assessment of the resistivity for large arrays of interconnect wires. Moreover, the accuracy of this method is reduced at decreasing wire dimensions. For narrow wires, the relative line width roughness (LWR) with respect to the line width is typically degraded, leading to issues caused by projection effects in TEM and a complex averaging behavior. 

For this reason, a method was proposed for the simultaneous determination of the cross-sectional area and the resistivity of interconnect wires using temperature-dependent measurements of the interconnect resistance \cite{SMJ_1994,LJP_1997,SVS_2001}. Since the method is based on the temperature coefficient of the resistivity (TCR), it is often referred to as the TCR method. However, the method was originally proposed for interconnect line widths of the order of 100\,nm and above and its applicability to sub-20-nm-wide interconnects is unclear. Two critical issues arise for scaled lines: first, the validity of the main approximation in the TCR method in presence of strong surface and grain boundary scattering remains to be clarified. In addition, lithographic and patterning limitations have led to increasing relative LWR, as mentioned above. Therefore, the validity of the TCR method for rough lines with an area-dependent resistivity needs to be assessed. Moreover, the volume fraction of interconnect lines occupied by barrier and liner layers increases steadily. While historically the contribution of the barrier and liner layer to the conductance could be neglected, this becomes more and more questionable in current and future technology nodes. It is therefore of interest to understand under which conditions the barrier/liner contribution can be safely neglected and the TCR method can be applied to extract an accurate cross-sectional area and resistivity of the center conductor.

The paper is organized as follows. First, the TCR method will be derived in Sec.~\ref{Sec_TCR} and its main approximation will be discussed. A semiclassical thin film resistivity model will then be introduced in Sec.~\ref{Sec_SC_TCR} that allows for an assessment of the thickness- and temperature-dependent resistivity as well as a qualitative discussion of the validity of the TCR method for scaled interconnects. Subsequently, the TCR method will be rederived in Sec.~\ref{Sec_rough_lines} for rough interconnect lines and finally, the influence of the presence of conductive barrier and liner layers will be discussed in Sec.~\ref{Sec_liner}. 

\section{The TCR method\label{Sec_TCR}}
 
The TCR method is based on the measurement of the temperature dependence of the resistance of an interconnect wire described by Pouillet's law 
\begin{equation}
\label{Pouillet}
R = \rho L/A. 
\end{equation} 
\noindent Here, $\rho$ is the resistivity of the conductor material in the wire whereas $L$ and $A$ are its length and cross-sectional area, respectively. The variation of the resistance as a function of temperature can then be expressed by the total derivative
\begin{align}
\label{Tot_diff}
\frac{dR}{dT} & = \frac{d}{dT} \left(\frac{\rho L}{A}\right) \nonumber \\
& = \frac{d \rho}{d T} \frac{L}{A} + \frac{d L}{d T} \frac{\rho}{A} - \frac{d A}{d T} \frac{\rho L}{A^2}.
\end{align}
\noindent The first term, $(d \rho/d T)(L/A)$, describes the resistance change due to the temperature dependence of the resistivity of the conductor, whereas the other two terms describe effects of thermal expansion. While thermal expansion has been historically included in the TCR method \cite{LJP_1997}, its contribution to the total resistance change is typically small ($\sim 1$\%) since relative resistivity changes $(1/\rho)(d\rho/dT)$ are of the order of $10^{-3}$\,K$^{-1}$ whereas thermal expansion coefficients are of the order of $10^{-5}$\,K$^{-1}$. Therefore, thermal expansion effects will be neglected in the following. In all formulas below, thermal expansion effects can however be easily reintroduced by simply subtracting them from ${dR}/{dT}$. Equation~(\ref{Tot_diff}) then reduces to
\begin{equation}
\label{TCR_1}
\frac{dR}{dT} = \frac{L}{A} \frac{d \rho}{d T}.
\end{equation}
\noindent Here, $\frac{d \rho}{d T}$ denotes the TCR of the \emph{interconnect line} under test. Generally, it should be considered as unknown. However, in certain cases, it can be assumed that the TCR of the interconnect line is \emph{equal to the bulk TCR}, \emph{i.e.}
\begin{equation}
\label{TCR_assumption}
\frac{d \rho}{d T} \equiv \left. \frac{d \rho}{d T}\right|_\mathrm{bulk},
\end{equation}
\noindent independently of the wire area $A$. This is the main assumption of the TCR method. Equation~(\ref{TCR_1}) then becomes
\begin{equation}
\label{TCR_2}
\frac{dR}{dT} = \frac{L}{A} \left. \frac{d \rho}{d T}\right|_\mathrm{bulk}.
\end{equation}

Equations~(\ref{Pouillet}) and (\ref{TCR_2}) lead to a set of two equations with two unknown variables, $\rho$ and $A$, as the length of the line is typically known. The system can be easily solved, resulting in 
\begin{align}
A & = L \left.\frac{d \rho}{d T}\right|_\mathrm{bulk} \left(\frac{dR}{dT}\right)^{-1} \label{TCR_Eq_1}; \\
\rho & = R \left.\frac{d \rho}{d T}\right|_\mathrm{bulk} \left(\frac{dR}{dT}\right)^{-1} \label{TCR_Eq_2}.
\end{align}
\noindent Hence, both resistivity and cross-sectional area can be determined simultaneously from measurements of the line resistance as well as its temperature dependence. To distinguish the area determined by the TCR method and that measured by physical characterization (\emph{e.g.} TEM), the former is typically called the \emph{electrical} area whereas the latter is called the \emph{physical} area. It is clear from the above derivation that a key requirement for the applicability of the TCR method is the validity of the approximation in Eq.~(\ref{TCR_assumption}).

\section{The TCR method in metallic nanowires\label{Sec_SC_TCR}}

The key approximation of the TCR method requires that the temperature dependence of the resistivity of nanowires is equal to that of the bulk metal in the measured temperature window [\emph{cf.} Eq.~(\ref{TCR_assumption})]. This is equivalent to the assumption that the combined effects that increase the resistivity with respect to the bulk are independent of temperature, \textit{i.e.} the resistivity can be written as 
\begin{equation}
\label{Matthiessen}
\rho(T) = \rho(0) + \rho_\mathrm{bulk}(T).
\end{equation}

For sufficiently pure nonmagnetic metallic systems, the bulk resistivity is determined by electron-phonon scattering following the Bloch--Gr\"uneisen law
\begin{equation}
\label{BG}
\rho\left( T\right) = \rho(0) + CT^5 \int_0^{\frac{\Theta_D}{T}} \frac{x^5}{\left(e^x - 1\right) \left(1 - e^{-x}\right)} dx,
\end{equation}
\noindent which is of the form of Eq.~(\ref{Matthiessen}). Here, $\rho(0)$ describes the residual resistivity at 0\,K, \emph{e.g.} due to scattering by impurities. $\Theta_D$ is the Debye temperature of the phonon system and $C$ a prefactor that can \emph{e.g.} be determined from the bulk room-temperature resistivity. In a sufficiently narrow temperature range around room temperature, the resistivity depends linearly on temperature and can be described by a temperature-independent TCR. 

When the TCR method was initially developed for large wire widths, deviations of interconnect and bulk resistivity were essentially only due to impurity scattering \cite{SMJ_1994,LJP_1997,SVS_2001}. In this case, Eq.~(\ref{Matthiessen}) can be considered as a special case of Matthiessen's rule since impurity scattering has typically been observed to be independent of temperature. Experimentally, Matthiessen's rule has been found to be valid in sufficiently pure metallic thick films, \textit{i.e.} for sufficiently large residual resistance ratios, leading to a temperature-independent $\rho(0)$ as in Eq.~(\ref{BG}). Therefore, the validity of Eq.~(\ref{TCR_assumption}) has been linked to the validity of Matthiessen's rule \cite{SMJ_1994,LJP_1997,SVS_2001}.

However, additional scattering mechanisms may strongly contribute to the resistivity of metallic nanowires with widths of 20\,nm and below, in particular surface roughness scattering as well as grain boundary scattering. Moreover, the physics of electron-phonon scattering may be altered in nanowires. It is therefore imperative to understand whether relations such as Eq.~(\ref{Matthiessen}) and more specifically Eq.~(\ref{BG}) also hold in scaled nanowires. It should be noted that this does not necessarily require the validity of Matthiessen's rule for all scattering contributions. The total scattering rate $1/\tau = f(\tau_\mathrm{SR}, \tau_\mathrm{GB}, \tau_\mathrm{i}, \tau_\mathrm{e-ph})$ can be formally written as a function of $\tau_\mathrm{SR}$, $\tau_\mathrm{GB}$, $\tau_\mathrm{i}$, and $\tau_\mathrm{e-ph}$, the scattering times due to surface roughness (SR), grain boundaries (GB), impurities (i) and the electron--phonon interaction (e--ph), respectively. A minimum requirement for the applicability of the approximation in Eq.~(\ref{TCR_assumption}) is then that the scattering rate can be written as $1/\tau = f(\tau_\mathrm{SS}, \tau_\mathrm{GB}, \tau_\mathrm{i}) + 1/\tau_\mathrm{e-ph}$ as long as $f(\tau_\mathrm{SS}, \tau_\mathrm{GB}, \tau_\mathrm{i})$ is independent of temperature and the electron--phonon interaction is bulk-like. In the following, this will be discussed in more detail.

\subsection{Temperature dependence of resistivity in nanowires and thin films}

The temperature dependence of the resistivity has been studied for several metallic nanowire \cite{BBR_2006, OYH_2008, YLL_2012, KHB_2015, CLX_2015} and thin film \cite{BH_1959,AWV_1984,V_1988, KSS_1997,SES_2002, KBS_2004,MZT_2010} systems where the cross-sectional area or the thickness were determined by physical characterization methods. Typically, an \emph{increase} of the TCR with respect to the bulk metals has been observed for nanowires. For thin films, the situation is more diverse and both bulk-like \cite{AWV_1984,V_1988}, as well as higher-than-bulk \cite{KSS_1997,SES_2002, KBS_2004,MZT_2010} and lower-than-bulk TCR values \cite{BH_1959} have been observed. Most measurements were consistent with the Bloch--Gr\"uneisen law in Eq.~(\ref{BG}) with an approximately linear temperature dependence of the resistivity $\propto T/\Theta_D$ above typically 100\,K, albeit in case of an increased TCR with lower apparent Debye temperatures. 

The underlying physics of the apparent reduction of the Debye temperature in metallic nanostructures has been discussed in terms of surface effects on the phonon band structure. Indeed, surface-sensitive phonon spectroscopy has demonstrated a reduction of the Debye temperature due to the presence of the surface and the resulting change in bonding geometry \cite{DKKG_1987,KYM_2000,CBP_2016}. In nanowires, the surface-to-volume ratio is large and therefore, the reduction of the surface Debye temperature may affect the Bloch--Gr\"uneisen resistivity. However, as already pointed out in Refs.~\cite{AWV_1984,V_1988} and discussed in more detail below, surface scattering as described in the original Fuchs-Sontheimer theory \cite{F_1938,S_1952} also results in an increase in TCR. These two contributions are difficult to separate in practice. Thermal conductivity measurements of metallic nanowires have frequently reported violations of the Wiedemann--Franz law or at least Lorenz numbers that deviate strongly from bulk values \cite{CLX_2015,OYH_2008,VRC_2009}. \emph{Ab initio} calculations of the electron--phonon interaction in nanowires have been reported but only for very small wires, much below realistic dimensions \cite{VG_2006,SLK_2012}. Thus, finite size effects on the electron--phonon interaction in metallic nanowires still need to be better understood to clarify this issue.

The observation of bulk-like TCR in thin films \cite{AWV_1984,V_1988} indicates that the central approximation in Eq.~(\ref{TCR_assumption}) does not necessary fail in metallic structures with reduced dimension. Moreover, the TCR method has led to electrical cross-sectional areas of Cu and Ru interconnect wires with cross-sectional areas down to $< 100$\,nm$^2$ that are reasonably consistent with physical cross-sectional areas deduced from TEM \cite{DKW_2017, DKG_2017}. However, the generalization of such results to any metallic interconnect system appears doubtful.

\subsection{Temperature-dependent semiclassical thin film resistivity modeling}

To shed some light on the issue, quantitative models for the temperature dependence of metallic nanowires or thin films are necessary. Currently, there is no well-established tractable model for the temperature dependence of the resistivity in nanowires. Semiclassical approaches \cite{DMV_2018} still require further quantitative verification while \emph{ab initio} methods have been successful to describe quantum effects in ultrasmall nanowires but are limited to low temperatures or treat finite temperatures via a heuristic mean free path \cite{ZSA_2008,JSP_2015,HBR_2016,L_2017}. The resistivity of thin films has typically been modeled by semiclassical models, such as the one originally derived by Mayadas and Shatzkes \cite{MS_1970} based on the original work of Fuchs and Sondheimer \cite{F_1938, S_1952} as well as quantum models for surface scattering \cite{TJM_1986, SXW_1995, ZG_2018}. Despite its many approximations, the semiclassical model of Mayadas and Shatzkes \cite{MS_1970} has the advantage to treat both surface roughness and grain boundary scattering simultaneously in a unified framework without the need to assume the validity of Matthiessen's rule. In this model, the resistivity of a metallic thin film with height $h$ is given by

\begin{align}
\label{MS}
\rho_{tf} = & \left[ \frac{1}{\rho_\mathrm{GB}} - \frac{6}{\pi\kappa\rho_0} \left(1-p\right) \int\displaylimits_0^{\pi/2} d\phi \int\limits_1^\infty dt \frac{\cos^2\phi}{H^2}\right. \nonumber \\
& \times \left.\left(\frac{1}{t^3} - \frac{1}{t^5}\right) \frac{1-e^{-\kappa Ht}}{1-pe^{-\kappa Ht}}\right]^{-1}
\end{align}
\noindent with $\rho_0$ the bulk resistivity of the metal, $\lambda$ the mean free path of the charge carriers in the metal, $g$ the linear intercept length between grain boundaries (the ``grain size''), $0 \le R \le 1$ the grain boundary reflection coefficient, $0 \le p \le 1$ a phenomenological parameter that describes the degree of specular scattering at the nanowire surfaces, and the following abbreviations
\begin{align}
\rho_\mathrm{GB} & = \rho_0 \left[ 1- 3\alpha /2 + 3\alpha^2 - 3\alpha^3\ln \left(1 + 1/\alpha\right)\right]^{-1}; \\
H & = 1 + \frac{\alpha}{\cos\phi\sqrt{1-1/t^2}}; \\
\alpha & = \frac{\lambda}{g}\frac{2R}{1-R}; \\
\kappa & = \frac{h}{\lambda}.
\end{align}

Equation~(\ref{MS}) does not depend explicitly on temperature but implicitly via the temperature dependence of $\lambda$ and $\rho_0$. However, it has been recognized that the product $\lambda \times \rho_0$ is a function of the morphology of the Fermi surface only and can therefore be assumed to be independent of temperature \cite{ME_2004,G_2016,DSM_2017,ZG_2017} when $k_BT \ll E_F$ with $k_BT$ the thermal and $E_F$ the Fermi energy. As this is true for relevant metals at room temperature, it can thus be assumed that 
\begin{equation}
\lambda(T) \times \rho_0(T) = C_0.
\end{equation}
\noindent For the bulk resistivity of sufficiently pure metals, the temperature-dependent resistivity is given by the Bloch--Gr\"uneisen law in Eq.~(\ref{BG}) with $\rho_0(0) \approx 0$. This allows for the calculation of the temperature dependence of $\lambda$ (as values of $C_0$ are available in the literature for numerous elemental metals \cite{G_2016,DSM_2017}) and therefore for the calculation of the temperature dependence of the resistivity. 

\begin{figure*}[t]
  \centering
  \includegraphics[width=120 mm]{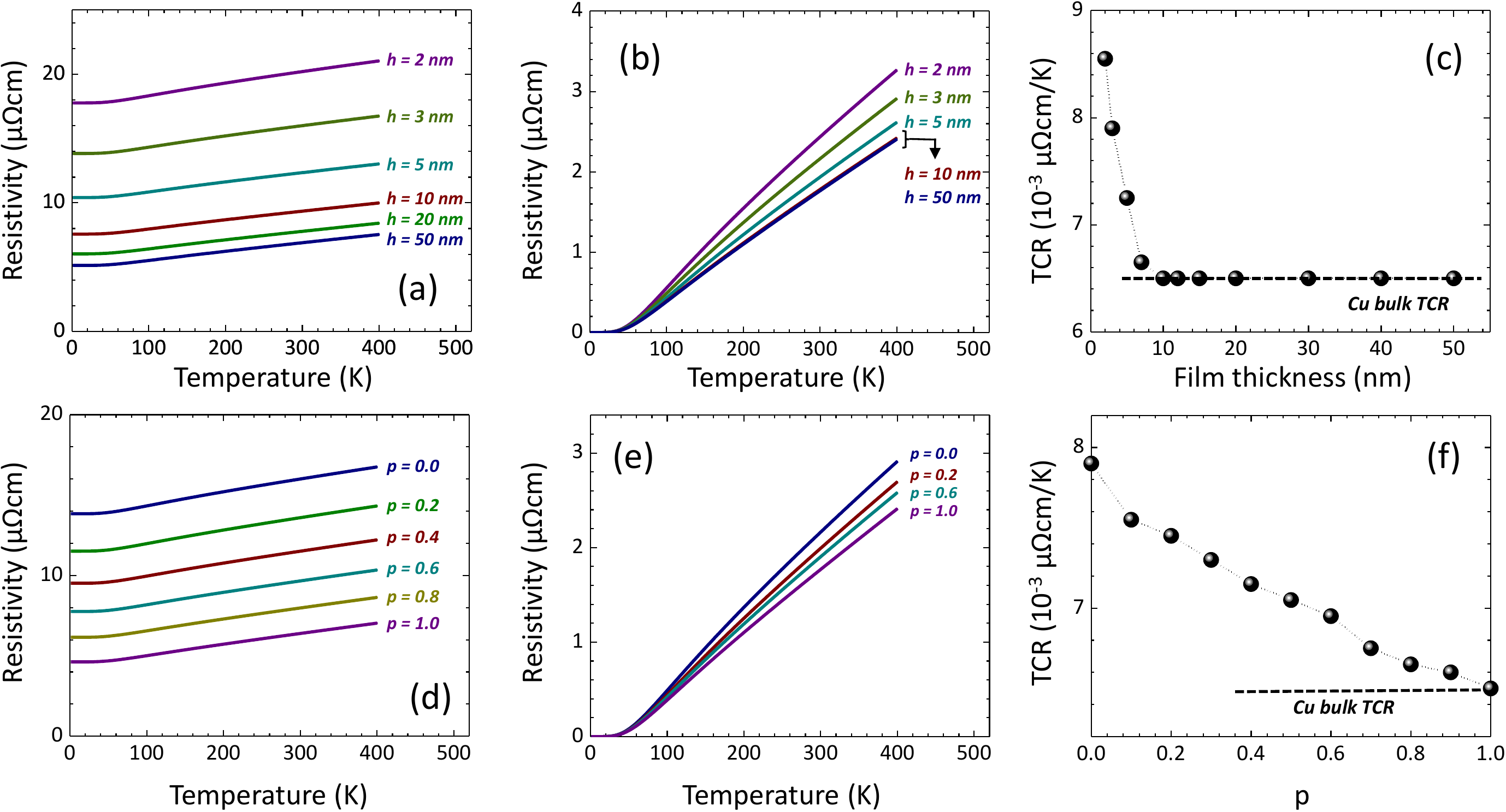}
\caption{\label{fig:CU_TCR} (a) Calculated temperature dependence of the resistivity of Cu thin films as a function of the film thickness using Eq.~(\ref{MS}) with $\lambda = 40.6$\,nm and $\rho_0 = 1.70$\,$\mu\Omega$cm at 300\,K, $\Theta_D = 310$\,K, $g = 10$\,nm, $p = 0$, $R = 0.2$. (b) Temperature-dependent part of the Cu thin film resistivity, $\rho(T) - \rho(0)$. (c) TCR at 300\,K calculated from the data in (a) and (b). (d) Calculated temperature dependence of the resistivity of a 3\,nm thick Cu film as a function of the surface scattering parameter $p$ using Eq.~(\ref{MS}). The same materials parameters as in (a) were used. (f) Temperature-dependent part of the 3\,nm Cu thin film resistivity in (d), $\rho(T) - \rho(0)$ and (f) TCR at 300\,K as a function of $p$.  For comparison, the bulk Ru TCR value is also shown [dashed lines in (c) and (f)] \cite{Bass_LB}.}
\end{figure*}

Marom and Eizenberg have derived analytical expressions of the TCR based on the Mayadas--Shatzkes model \cite{ME_2004}. In practice, it is more convenient to directly calculate numerically the temperature dependence of the resistivity via Eqs.~(\ref{BG}) and (\ref{MS}) and obtain the TCR by numerical differentiation. Calculations of the temperature dependence of the resistivity for Cu ($\lambda = 40.6$\,nm and $\rho_0 = 1.70$\,$\mu\Omega$cm at 300\,K, $\Theta_D = 310$\,K, $g = 10$\,nm, $p = 0$, $R = 0.2$) as a function of the film thickness $h$ are shown in Fig.~\ref{fig:CU_TCR}(a--c). The chosen $p$ and $R$ values are typically obtained from experimental thickness dependences of the resistivity \cite{JBT_2009} with the Mayadas--Shatzkes model. The total calculated resistivity in Fig.~\ref{fig:CU_TCR}(a) indicates a strong increase of the resistivity with decreasing thickness, as expected. By contrast, the temperature dependent part of the resistivity, $\rho(T) - \rho(0)$, in Fig. \ref{fig:CU_TCR}(b) and the extracted TCR value at 300\,K in \ref{fig:CU_TCR}(c) are bulk-like \cite{Bass_LB} for thicker films but increase for thicknesses below 10\,nm. To elucidate this behavior further, the Cu thin film resistivity was calculated for a thickness of $h = 3$\,nm as a function of the surface roughness scattering parameter $p$. The total calculated resistivity in Fig.~\ref{fig:CU_TCR}(d) increases with decreasing $p$ since this enhances the contribution of surface roughness scattering. The temperature dependent part of the resistivity, $\rho(T) - \rho(0)$, in Fig.~\ref{fig:CU_TCR}(e) and the extracted TCR value at 300\,K in \ref{fig:CU_TCR}(f) also show a strong dependence on $p$. For specular $p \sim 1$, the TCR is bulk-like and increases rapidly with decreasing $p$. Moreover, additional simulations (not shown) indicate that the TCR decreases again towards the bulk value when the grain boundary scattering is enhanced by increasing $R$ or decreasing $g$. This indicates that it is essentially the ratio of surface and grain boundary scattering that determines the TCR, which can be attributed to the ``renormalization'' of the (effective) mean free path due to grain boundary scattering, as discussed in detail in Ref.~\cite{MS_1970}.

\begin{figure}[t]
  \centering
  \includegraphics[width=80 mm]{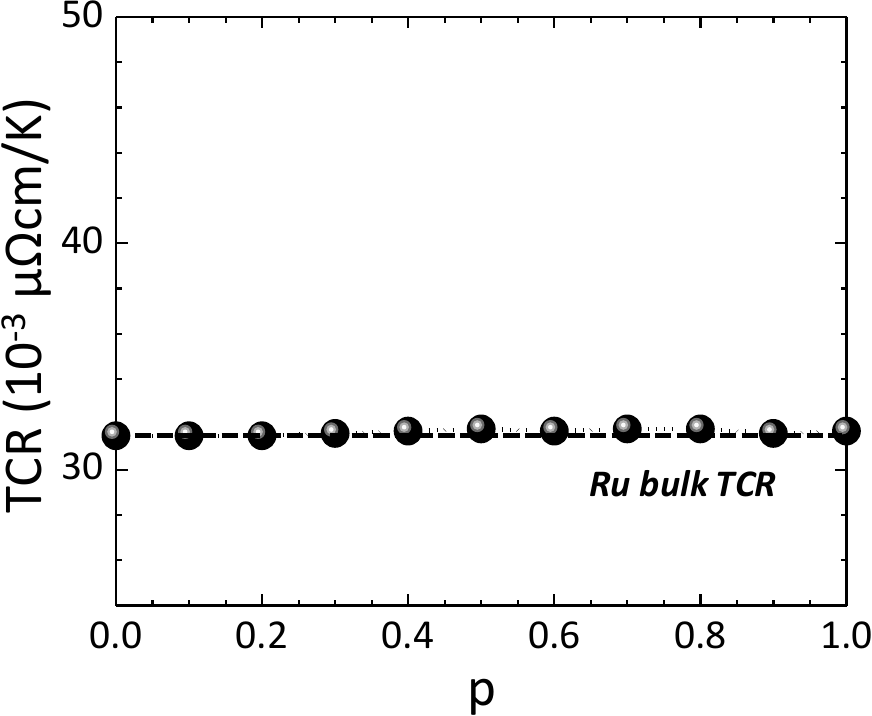}
\caption{\label{fig:RU_TCR} (a) Calculated room-temperature TCR of a 3\,nm thick Ru film for different values of the surface scattering parameter $p$ following Eq.~(\ref{MS}) with $\lambda = 6.6$\,nm and $\rho_0 = 7.6$\,$\mu\Omega$cm at 300\,K, $\Theta_D = 420$\,K, $g = 10$\,nm, $R = 0.5$. The thin film TCR is equal to the bulk TCR (dashed line) for all values of $p$. }
\end{figure}

Figure~\ref{fig:RU_TCR} show the calculated TCR at 300\,K for Ru ($\lambda = 6.6$\,nm and $\rho_0 = 7.6$\, $\Theta_D = 420$\,K, $\mu\Omega$cm at 300\,K, $g = h$, $R = 0.5$). As for Cu, the experimentally determined $R$ value for Ru is used \cite{DSM_2017}. In contrast to Cu, $p$ has no influence on the TCR for the chosen set of parameters. This can be linked to the shorter mean free path $\lambda$ of Ru as well as the higher $R$, which increases the relative importance of grain boundary scattering and renders surface roughness scattering negligible. These results are consistent with the experiments and the discussion in Ref.~\cite{DSM_2017}, which has also reported that surface scattering in Ru thin films is negligible even for $p = 0$. The results are also consistent with initial experiments on Ru thin films indicating bulk TCR values for Ru films as thin as 3\,nm (not shown). This will be the topic of a forthcoming publication \cite{SFM_unpub}.

The data indicate that the TCR within such semiclassical models is influenced mainly by the ratio of surface roughness and grain boundary scattering. When grain boundary scattering dominates, the TCR is bulk-like. However, when the effect of surface scattering becomes non-negligible, the TCR starts to deviate from its bulk value. Since the Mayadas--Shatzkes model does not obey Matthiessen's rule for surface roughness and grain boundary scattering, the individual contributions cannot be regarded separately and the relative importance of surface scattering needs to be deduced from an evaluation of Eq.~(\ref{MS}), \emph{e.g.} by comparing resistivities calculated for $p = 1$ and $p = 0$. Note that the value of $p$ as such does not determine the importance of surface roughness scattering. In particular, $p = 0$ does not necessarily indicate strong surface roughness scattering. For typical thin films with $g \approx h$, grain boundary scattering is generally dominant for Cu and Ru even with $p= 0$ \cite{DSM_2017} and calculated TCR values are within a few \% of the bulk value even for ultrathin films of 3\,nm thickness. However, attempts to increase the grain size, \emph{e.g.} by thermal annealing (leading to recrystallization and/or grain growth), may change this picture. It should also be mentioned that for ultrathin films or ultranarrow wires, the surface roughness may become comparable to the film thickness or wire width at the limit of film closure or wire continuity, and the used model might lose its applicability \cite{ZZP_2018}. In general, alternative metals with short mean free path and high grain boundary reflection coefficient $R$ (which is correlated with higher cohesive energy \cite{ZL_2010} and thus improved expected electromigration resistance) should however show less tendency for an increase in TCR than Cu. A more detailed discussion can be found in Ref.~\cite{ME_2004}.

The above results indicate that the ratio of surface to grain boundary scattering determines the deviation of the thin film TCR from bulk values. One might assume that a similar relation also holds qualitatively for nanowires. Due to the larger surface-to-volume ratio, surface scattering can be expected to have a larger impact for nanowires than for thin films, as qualitatively argued in Ref.~\cite{DMV_2018}. In absence of a comprehensive quantitative model for temperature-dependent nanowire resistivities, the TCR measured for thin films might be used as a proxy for the expected nanowire TCR. When the TCR of thin films is strongly enhanced with respect to the bulk value, one can assume that the TCR of nanowires might also deviate. By contrast, a bulk-like thin film TCR might not necessarily indicate a bulk-like nanowire TCR. A potential improvement could the usage of measured thin film TCR values to determine cross-sectional area and resistivity of nanowires of the same material. However, in absence of a predictive model for nanowire resistivity (and the rather simplistic nature of the semiclassical thin film models), it is currently not possible to verify the accuracy of this approach.

Finally, in its derivation, the Mayadas-Shatzkes model assumes the validity of Matthiessen's rule for grain boundary and phonon scatting (but not for surface roughness scattering) \cite{MS_1970}. As argued above, this is a reasonable assumption for sufficiently pure films. However, in presence of strong disorder, Matthiessen's rule has been reported to be violated, leading to a much weaker temperature dependence of the resistivity and even to zero or negative TCR values \cite{M_1973,T_1986}. This suggests that the low thin film TCR values reported in Ref.~\cite{BH_1959} might be due to large disorder in the films. Such effects cannot be described within the Mayadas-Shatzkes model and their quantification is rather difficult, leading to further potential complications of the determination of TCR values of thin films and nanowires. The experimental reports mentioned above however suggest that the disorder in current nanowire (and interconnect) structures is sufficiently low so that disorder effects might be neglected.

\section{The TCR method for a rough interconnect line\label{Sec_rough_lines}}

\subsection{Model derivation and discussion\label{SubSec_rough_lines}}

An additional complication that affects the application of the TCR method to scaled interconnect lines stems from an increased LWR due to lithography and patterning limitations. The derivation of the TCR method in Eqs.~(\ref{TCR_Eq_1}) and (\ref{TCR_Eq_2}) relies on the assumption of a single uniform cross-sectional area $A$ of the interconnect wire. However, in scaled interconnects, this assumption ceases to be obviously justified. This is even more the case for experimental patterning schemes \cite{DKG_2017, DKW_2017}. In such cases, the $3\sigma$ LWR may not be negligible anymore with respect to the average line width $\left< w\right>$. Therefore, it is of interest to redevelop the TCR equations by assuming a cross-sectional area that varies along the wire.

A rough interconnect line can be described by a series of small differential resistive elements of length $d\ell$ and resistance $dR$ so that the total resistance of the line with length $L$ is given by $R = \int_0^L dR$. Assuming that each element can be described by Pouillet's law in Eq.~(\ref{Pouillet}), $dR = \left( \rho[A(\ell)]/A(\ell)\right) d\ell$. Here, $A(\ell )$ is the cross-sectional area of the element and $\rho[A(\ell)]$ its resistivity. Note that the resistivity depends in general on the cross-sectional area of the element due to finite size effects as discussed above. The resistance of the line can then be written as
\begin{equation}\label{Resistance}
R = \int_0^L \frac{\rho[A(\ell)]}{A(\ell )} d\ell .
\end{equation}

Neglecting thermal expansion effects, the total derivative of the resistance with respect to the temperature is given by
\begin{equation}
\frac{dR}{dT} = \int_0^L \frac{d\rho[A(\ell)]}{dT}\frac{1}{A(\ell )} d\ell .
\end{equation}
\noindent Within the approximation of Eq.~(\ref{TCR_assumption}), $d\rho[A(\ell)]/dT$ is independent of the cross-sectional area and thus
\begin{align}
\label{A_harmon}
\frac{dR}{dT} & = \left. \frac{d\rho}{dT}\right|_\mathrm{bulk} \int_0^L \frac{1}{A(\ell )} d\ell \nonumber \\
& = \left. \frac{d\rho}{dT}\right|_\mathrm{bulk} \frac{L}{\left< A \right >_H},
\end{align}
\noindent with $\left< A\right>_H = L/\left(\int_0^L 1/A(\ell )d\ell\right)$ the \textit{harmonic} mean of the area function $A(\ell)$. 

This shows that the TCR method yields the harmonic mean of the cross-sectional area for a rough line. The harmonic mean is always smaller than or equal to the arithmetic mean $\left<A\right>_A$; in the limit of a uniform area, harmonic and arithmetic means become equal, \emph{i.e.} $\lim_{\sigma^2\to 0} (\left<A\right>_A - \left<A\right>_H) = 0$ with $\sigma^2$ the variance of the area distribution.

Equation~(\ref{Resistance}) cannot be evaluated for a general $\rho[A(\ell)]$ and therefore the resistivity cannot be determined in a straightforward way as in Eq.~(\ref{TCR_Eq_2}). However, a Taylor expansion of $\rho[A(\ell)]$ around $\left< A \right >_H$ can be used to approximate the area dependence of the resistivity. Thus, 
\begin{align}
\rho[A(\ell)] & = \rho\left(\left<A\right>_H \right) + \left.\frac{d\rho}{dA}\right|_{\left<A\right>_H} \hspace{-0.4cm} \left(A(\ell) - \left<A\right>_H\right) \nonumber \\
& \hspace{0.5cm} + \frac{1}{2}\left.\frac{d^2\rho}{dA^2}\right|_{\left<A\right>_H} \hspace{-0.4cm} \left(A(\ell) - \left<A\right>_H\right)^2 + \ldots \nonumber \\
& \equiv \left<\rho\right> + \xi\left(A(\ell) - \left<A\right>_H\right) \nonumber \\
& \hspace{0.5cm} + \frac{1}{2}\eta\left(A(\ell) - \left<A\right>_H\right)^2 + \ldots
\end{align}
\noindent with $\left<\rho\right>$ the resistivity at the harmonic mean of $A(\ell )$. $\xi$ and $\eta$ denote the first and second derivatives of $\rho(A)$ at $\left<A\right>_H$, respectively. In a first-order approximation, the resistance in Eq.~(\ref{Resistance}) then becomes
\begin{align}
R & = \int_0^L \frac{\rho(\ell )}{A(\ell )} d\ell \approx \int_0^L \frac{\left<\rho\right> + \xi \left(A(\ell) - \left<A\right>_H\right)}{A(\ell )} d\ell \nonumber \\
& = \frac{\left<\rho\right> L}{\left<A\right>_H} + L\xi - \xi\left<A\right>_H \int_0^L \frac{1}{A(\ell )} d\ell \nonumber \\
& = \frac{\left<\rho\right> L}{\left<A\right>_H} \label{Res_rough}
\end{align}
\noindent Solving the system of Eqs.~(\ref{A_harmon}) and (\ref{Res_rough}) leads then to the first-order TCR equations for rough lines, analogously to Eqs.~(\ref{TCR_Eq_1}) and (\ref{TCR_Eq_2}):
\begin{align}
\left<A\right>_H & = L \left.\frac{d \rho}{d T}\right|_\mathrm{bulk} \left(\frac{dR}{dT}\right)^{-1}; \label{Rough_TCR_Eq_1} \\
\rho\left(\left<A\right>_H\right) & = R \left.\frac{d \rho}{d T}\right|_\mathrm{bulk} \left(\frac{dR}{dT}\right)^{-1}. \label{Rough_TCR_Eq_2}
\end{align}
\noindent The TCR method yields thus the harmonic mean of the cross-sectional area $\left<A\right>_H$ and to first order the resistivity at $\left<A\right>_H$. Therefore, the resistivity \textit{vs.} area curve extracted by the TCR method from a set of rough wires with different $\left<A\right>_H$ represent quantitatively (to first order) the area dependence of the resistivity $\rho(A)$.

The accuracy of the first-order Taylor expansion can be tested by calculating the second-order correction to Eq.~(\ref{Res_rough}), leading to
\begin{align}
\Delta R & = \frac{\eta}{2} \int_0^L \frac{\left(A(\ell) - \left<A\right>_H\right)^2}{A(\ell)} d\ell \nonumber \\
& = \frac{\eta}{2} \left(L\left<A\right>_A  - 2L\left<A\right>_H + L \left<A\right>_H \right) \nonumber \\
&  = \frac{L\eta}{2}\left(\left<A\right>_A - \left<A\right>_H \right).
\end{align}
\noindent Thus, the correction is proportional to the second derivative of $\rho(A)$ as well as the difference between arithmetic and harmonic means of $A(\ell)$. Note that for uniform lines, $\left<A\right>_A = \left<A\right>_H \equiv A$ as well as $\Delta R = 0$ and Eqs.~(\ref{Rough_TCR_Eq_1}) and (\ref{Rough_TCR_Eq_2}) turn into Eqs.~(\ref{TCR_Eq_1}) and (\ref{TCR_Eq_2}).

The second-order correction $\Delta R$ cannot be evaluated for a general $A(\ell)$. However, since $\left<A\right>_A \geq \left<A\right>_H \ge 0$, an upper limit is given by
\begin{equation}
\label{UL}
\Delta R \leq \frac{L\eta}{2}\left<A\right>_A 
\end{equation}
\noindent Practically, the upper limit of the second-order correction could \emph{e.g.} be estimated by the apparent physical cross-sectional area (\emph{e.g.} measured by TEM) as an estimator of $\left<A\right>_A$ and the second derivative of the obtained $\rho(A)$ curve. For small Cu and Ru wires, $\eta_\mathrm{Cu} \sim 10^{-4}$\,$\mu\Omega$cm/nm$^4$ and $\eta_\mathrm{Ru} \sim 10^{-5}$\,$\mu\Omega$cm/nm$^4$, respectively. For a very rough wire with an arithmetic average area of 100\,nm$^2$ (and a harmonic average area of $\sim 0$\,nm$^2$), the correction is thus below 10\,$\Omega/\mu$m  for Ru and 100\,$\Omega/\mu$m for Cu. Typical line resistances for 100\,nm$^2$ cross-sectional area are above $1$\,k$\Omega/\mu$m and thus the corrections are small to negligible even in such an extreme case. The corrections are not expected to depend strongly on the cross-sectional area as $\eta$ decreases rapidly with increasing cross-sectional area or line width. As shown in the next section, $\left<A\right>_A - \left<A\right>_H \ll \left<A\right>_A$ for realistic LWRs and thus the real values of $\Delta R$ will be much smaller. Therefore, second-order corrections can typically be neglected.

\subsection{Case study: truncated Gaussian LWR statistics}

The variability of the dimensions of an interconnect wire is typically described by the rms LWR as well as a lateral correlation length (or the roughness power spectrum). As the correlation length is not expected to affect much the resistivity of the conductor as long as Pouillet's law remains applicable, it is mainly of interest to understand the dependence of the electrical cross-sectional area determined by TCR, $\left<A\right>_H$, on the LWR. 

Let us consider a rectangular interconnect line with width $w$, height $h$, and area $A = w \times h$. The width $w$ is treated as a stochastic variable characterized by a probability function that can be described by an arithmetic mean $\left< w\right>$ and a variance $\sigma_w^2$. For simplicity, it is assumed that the line height $h$ does not vary along the line and thus $\left<A\right>_A = \left<w\right>\times h$ and $\sigma^2_A = \sigma^2_w \times h^2$. The line height variation with an average height of $\left<h\right>$ and a variance $\sigma^2_h$ can however be easily included in the model using $\left<A\right>_A = \left<w\right>\times \left<h\right>$ and $\sigma^2_A = \sigma^2_w \times \left<h\right>^2 + \sigma^2_h \times \left<w\right>^2$. Experimentally, the line width has been found to follow a Gaussian distribution \cite{LLE_2005}. The probability function for the cross-sectional area is then
\begin{equation} 
P(A) = \frac{1}{\sqrt{2\pi\sigma^2_A}}\exp\left( {-\frac{\left(A - \left<A\right>_A\right)^2}{2\sigma^2_A}}\right)
\end{equation}
\noindent and the harmonic average of the area can then be calculated by
\begin{equation}
\label{Gauss_harm_1}
\left< A\right>_H = \left( \int_{-\infty}^\infty \frac{P(A)}{A} dA\right)^{-1}.
\end{equation}

From a mathematical viewpoint, the usage of Gaussian statistics is somewhat problematic since a Gaussian has a finite probability at $A = 0$ and therefore the harmonic average of the area is always zero. In practice however, for interconnects with high yield, Gaussian statistics can be assumed to be obeyed only in a certain window around the mean line width with a rapid falloff to zero probability outside. This means that Eq.~(\ref{Gauss_harm_1}) can be approximately written as 
\begin{equation}
\label{Gauss_harm_2}
\left< A\right>_H = \left( \int_{\left<A\right>_A-c}^{\left<A\right>_A+c} \frac{P(A)}{A} dA\right)^{-1},
\end{equation}
\noindent with a suitably chosen cutoff $c$. If $\left<A\right>_A \gg \sigma_A$, the divergence of $P(A)/A$ at $A = 0$ is slow and $\left< A\right>_H$ does not depend critically on the choice of $c$.

\begin{figure}[t]
  \centering
  \includegraphics[width=80 mm]{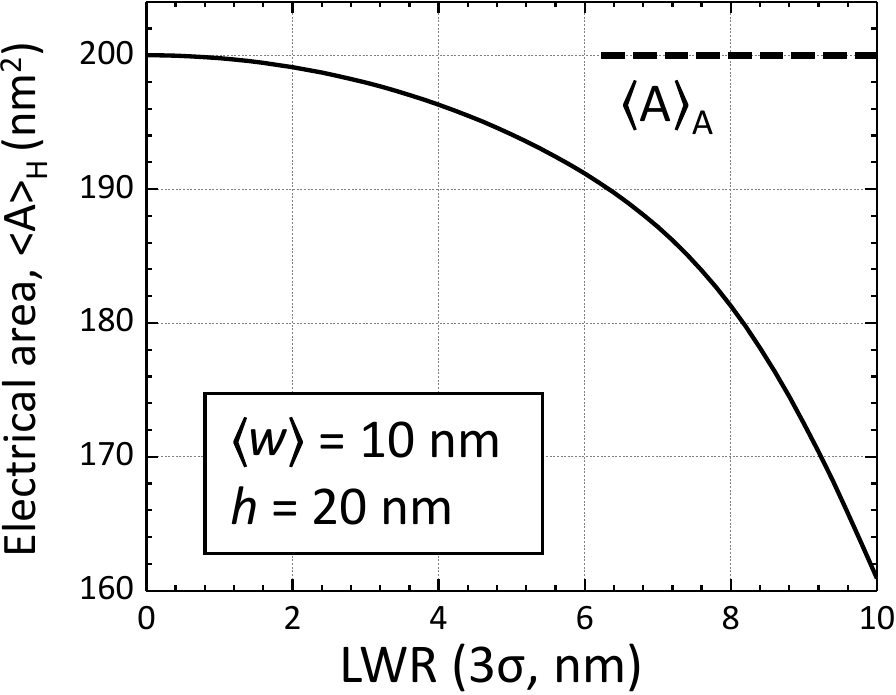}
\caption{\label{fig:Harm_Ave} Harmonic area average, $\left< A\right>_H$, assuming a truncated Gaussian distribution for the interconnect line width as a function of the 3$\sigma$ LWR. The mean line width $\left<w\right>$ was 10\,nm and the line height $h$ was fixed at 20\,nm, as discussed in the text, \emph{i.e.} the mean cross-sectional area (arithmetic mean) was 200\,nm$^2$ (dashed line).}
\end{figure}

Figure~\ref{fig:Harm_Ave} shows the calculated value of $\left< A\right>_H$ for a line with an average width of $\left<w\right> = 10$\,nm and a fixed height of $h = 20$\,nm as a function of the $3\sigma$ LWR under the above assumptions. Here, the cutoff was chosen as $\left<A\right>_A-c = 0.1$\,nm$^2$. As expected, the choice of the cutoff close to zero had negligible influence on $\left<A\right>_H$. 

One can see that an increasing $3\sigma$ LWR leads to a reduction of $\left< A\right>_H$ with respect to $\left< A\right>_A = 200$\,nm$^2$. However, rather large values of $3\sigma \approx 8$\,nm are required for a 10\% reduction of $\left< A\right>_H$ with respect to $\left< A\right>_A$. In real interconnects, variations of the line height will lead to larger area variances and to larger deviations, as discussed above. Nonetheless, the above case study shows that the TCR method can be expected to lead to accurate mean line areas if the $3\sigma$ LWR is much smaller than the average line width, which is typically the case for interconnects with high yield.

\section{The TCR method for an interconnect line with a conductive liner\label{Sec_liner}}

In scaled interconnects, the volume available for the center conductor (commonly Cu) is increasingly reduced because barrier and liner layers are difficult to scale. In former technology nodes, TaN/Ta was used as the barrier/liner combination occupying typically a volume fraction of less than 10\%{}. With a Cu resistivity close to the bulk value and much larger Ta and TaN resistivities, the contribution of the liner and barrier to the line conductivity was negligible. In this case, barrier and liner could be treated as insulators from an electrical point of view and the TCR method could be used to extract area and resistivity of the center conductor.

However, in future (Cu-based) technology nodes, the volume fraction occupied by barrier and liner(s) may reach several 10\%{} at line widths of the order of 10\,nm. In addition, lower resistivity liners, such as Co and Ru have been introduced. Therefore, the parallel conductance due to barriers and liners may not be negligible and may affect the TCR analysis of the interconnect wire. It is evident that the extraction of four unknown parameters, \textit{i.e.} the resistivities of the center conductor, $\rho_\mathrm{cc}$, and the liner, $\rho_\mathrm{li}$, as well as their (electrical) areas, $A_\mathrm{cc}$ and $A_\mathrm{li}$, is not possible from the two available measurements, $R$ and $dR/dT$. Therefore, an accurate TCR analysis of interconnects with a significant contribution of the barrier/liner combination to the conductance is not possible without further knowledge, \emph{e.g.} of the resistivity and the area of the liner. However, the conditions, under which the assumption of an ``insulating'' barrier/liner is valid and the TCR method yields accurate values for the resistivity and area of the center conductor can be discussed. Within such an approach, when approximate values for resistance, resistivity, and area ratios of center conductor and barrier/liner are available, approximate correction factors can be calculated and errors due to the presence of barrier and liner can be estimated.

The resistance of an interconnect line consisting of two parallel conductors, \textit{i.e.} the center conductor with resistance $R_\mathrm{cc}$ and a barrier/liner with resistance $R_\mathrm{li}$ is given by

\begin{equation} \label{Eq_paralel_R}
R = \left( \frac{1}{R_\mathrm{cc}} + \frac{1}{R_\mathrm{li}} \right)^{-1} = \frac{R_\mathrm{cc}R_\mathrm{li}}{R_\mathrm{cc} + R_\mathrm{li}}.
\end{equation}

\noindent The total derivative of the resistance with respect to the temperature is then

\begin{align}
\frac{dR}{dT} & = \frac{\partial R}{\partial R_\mathrm{cc}} \frac{d R_\mathrm{cc}}{dT} + \frac{\partial R}{\partial R_\mathrm{li}} \frac{d R_\mathrm{li}}{dT} \nonumber \\
& = \frac{R^2_\mathrm{li}}{\left( R_\mathrm{cc} + R_\mathrm{li} \right)^2}\frac{d R_\mathrm{cc}}{dT} + \frac{R^2_\mathrm{cc}}{\left( R_\mathrm{cc} + R_\mathrm{li} \right)^2}\frac{d R_\mathrm{li}}{dT} \nonumber \\
& = \frac{R^2}{R^2_\mathrm{cc}}\frac{d R_\mathrm{cc}}{dT} + \frac{R^2}{R^2_\mathrm{li}}\frac{d R_\mathrm{li}}{dT} .
\label{DT_parallel_R}
\end{align}

In principle, when all necessary information about the liner properties ($A_\mathrm{li}$, $\rho_\mathrm{li}$, $d\rho_\mathrm{li}/dT$) is available, Eqs.~(\ref{Eq_paralel_R}) and (\ref{DT_parallel_R}) can be solved to deduce $A_\mathrm{cc}$ and $\rho_\mathrm{cc}$ in the same TCR approximation as above. However, this will be rarely the case in practice and therefore explicit formulae for $A_\mathrm{cc}$ and $\rho_\mathrm{cc}$ are only of limited interest. By contrast, for current interconnects, the contribution of the liner to the total line conductance will still be rather small. In such cases, one can then ask the question, under which conditions the liner contribution is sufficiently small to lead to acceptable errors in the extraction of $A_\mathrm{cc}$ and $\rho_\mathrm{cc}$ when the parallel conductance of barrier/liner is neglected. To this aim, $R_\mathrm{li}$ can be treated as a ``perturbation'', leading to appropriate correction factors for the TCR Eqs. (\ref{TCR_Eq_1}) and (\ref{TCR_Eq_2}). Although the correction factors may not be accurately quantifiable in practice, approximate values can be determined when the liner resistance can be \textit{estimated} and the accuracy of the TCR method can be semi-quantitatively assessed. 

For a weak liner contribution, it can be assumed that $R_\mathrm{cc} \ll R_\mathrm{li}$ and all quadratic terms in $R_\mathrm{cc}/R_\mathrm{li}$ can be neglected. Thus, the total differential in Eq.~(\ref{DT_parallel_R}) becomes
\begin{align}
\frac{dR}{dT} & = \frac{R^2_\mathrm{li}}{\left( R_\mathrm{cc} + R_\mathrm{li} \right)^2}\frac{d R_\mathrm{cc}}{dT} + \frac{R^2_\mathrm{cc}}{\left( R_\mathrm{cc} + R_\mathrm{li} \right)^2}\frac{d R_\mathrm{li}}{dT} \label{Eq_diff_R_liner1}\\
& \approx R_\mathrm{cc}\frac{1}{\left( 1 + R_\mathrm{cc}/R_\mathrm{li} \right)^2} \frac{1}{\rho_\mathrm{cc}}\frac{d\rho_\mathrm{cc}}{dT}  \nonumber \\
& \hspace{0.5cm} + \frac{R^2_\mathrm{cc}}{R_\mathrm{li}}\frac{1}{\left( 1 + R_\mathrm{cc}/R_\mathrm{li} \right)^2} \frac{1}{\rho_\mathrm{li}}\frac{d\rho_\mathrm{li}}{dT}. \label{Eq_diff_R_liner2}
\end{align}

\noindent A further quantitative analysis of the above equations to obtain electrical areas of the center conductor and/or the liner requires the knowledge of the TCR of the liner, $dR_\mathrm{li}/dT$ or alternatively $d\rho_\mathrm{li}/dT$. Some quantitative insight in the effect of the liner can be gained from two different approximations. For many pure bulk metals, values of $(1/\rho )(d\rho /dT )$ are of the same order of magnitude. For example, $(1/\rho )(d\rho /dT )$ at 0\,\deg C is 0.0043\,K$^{-1}$ for Cu, 0.0056\,K$^{-1}$ for Co, or 0.0045\,K$^{-1}$ for Ru \cite{Bass_LB}. Therefore, one can assume that the relative temperature derivatives of the resistivities of center conductor and liner are equal, \textit{i.e.} $(1/\rho_\mathrm{cc})(d\rho_\mathrm{cc}/dT) = (1/\rho_\mathrm{li})(d\rho_\mathrm{li}/dT)$. In addition, for ``highly defective'' (or impure) layers, one can often assume $d\rho /dT \sim 0$ \cite{BH_1959} since Matthiessen's rule is violated in this case. This amounts to neglecting the second term $\propto d\rho_\mathrm{li}/dT$ in Eq.~(\ref{Eq_diff_R_liner2}).

Assuming $d\rho_\mathrm{li}/dT = 0$, Eq.~(\ref{Eq_diff_R_liner1}) can be combined with Pouillet's law in Eq.~(\ref{Pouillet}) to obtain the electrical area of the center conductor, leading to

\begin{equation}
A_\mathrm{cc} = \alpha L \left.\frac{d \rho_\mathrm{cc}}{d T}\right|_\mathrm{bulk} \left(\frac{dR}{dT}\right)^{-1}, \label{Liner_TCR_Eq_1}
\end{equation}

\noindent with the correction factor due to the contribution due to the liner

\begin{equation}
\alpha = \frac{1}{\left( 1 + R_\mathrm{cc}/R_\mathrm{li} \right)^2}.
\label{Eq_alpha_full_1}
\end{equation}

\noindent For small $R_\mathrm{cc}/R_\mathrm{li} \lesssim 0.1$, the relation can be approximated as $\alpha\approx 1 - 2R_\mathrm{cc}/R_\mathrm{li}$.

Alternatively, one can assume that the relative temperature derivatives of the resistivities of center conductor and liner are equal, \textit{i.e.} $(1/\rho_\mathrm{cc})(d\rho_\mathrm{cc}/dT) = (1/\rho_\mathrm{li})(d\rho_\mathrm{li}/dT)$. This leads to the same equation of the electrical area of the center conductor as in Eq.~(\ref{Liner_TCR_Eq_1}), albeit with a correction factor of

\begin{equation}
\alpha = \frac{1 + \frac{R^2_\mathrm{cc}}{R^2_\mathrm{li}}\frac{A_\mathrm{cc}}{A_\mathrm{li}}}{\left( 1 + R_\mathrm{cc}/R_\mathrm{li} \right)^2}.
\label{Eq_alpha_full_2}
\end{equation}

\noindent Here, $A_\mathrm{cc}$ and $A_\mathrm{li}$ are the cross-sectional areas of the center conductor and the liner, respectively.

\begin{figure}[t]
  \centering
  \includegraphics[width=80 mm]{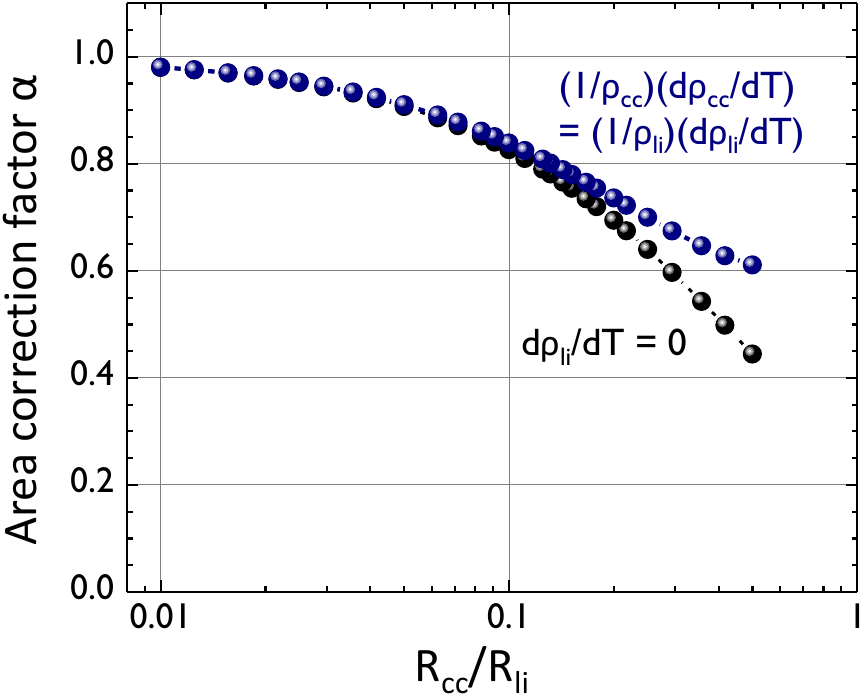}
\caption{\label{Fig_alpha} Electric area correction factor $\alpha$ assuming $d\rho_\mathrm{li}/dT = 0$ (black dots) and assuming equal relative TCR values of center conductor and liner, \emph{i.e.} $(1/\rho_\mathrm{cc})(d\rho_\mathrm{cc}/dT) = (1/\rho_\mathrm{li})(d\rho_\mathrm{li}/dT)$ (blue dots).}
\end{figure}

In all relevant cases, when the liner resistivity is smaller than that of the center conductor, $\alpha \le 1$. Therefore, ignoring the contribution of the liner to the total line resistance, \emph{i.e.} assuming $\alpha = 1$, leads to an \emph{overestimation} of the area of the center conductor. Of course, the correction factor $\alpha$ cannot be calculated in the general case as $R_\mathrm{cc}$ and $R_\mathrm{li}$ are \textit{a priori} unknown. However, Eq.~(\ref{Liner_TCR_Eq_1}) can be used to estimate the error of the electrical area when a reasonable estimate of the liner and center conductor resistances can be obtained (consistent with the total measured resistance). An example plot of the correction factor $\alpha$ as a function of $R_\mathrm{cc}/R_\mathrm{li}$ is shown in Fig.~\ref{Fig_alpha}. The figure shows the correction factors derived from Eqs.~(\ref{Eq_alpha_full_1}) and (\ref{Eq_alpha_full_2}) for $A_\mathrm{cc}/A_\mathrm{li} = 1.5$. It is clear from Eqs.~(\ref{Eq_alpha_full_1}) and (\ref{Eq_alpha_full_2}) that absolute values of $R_\mathrm{cc}$ and $R_\mathrm{li}$ do not affect $\alpha$, which depends only on the ratio. The graph shows that corrections start to become considerable (of the order of 20\%) when the liner contribution approaches 10\% of the total conductance of the line, rather independent of the approximation of Eqs.~(\ref{Eq_diff_R_liner1}) and (\ref{Eq_diff_R_liner2}). 

In the second step, the resistivity of the center conductor will be assessed in presence of the barrier/liner. Eq.~(\ref{Eq_paralel_R}) can be approximated to first order as

\begin{equation}
R = \frac{R_\mathrm{cc}R_\mathrm{li}}{R_\mathrm{cc} + R_\mathrm{li}} \approx R_\mathrm{cc} \left( 1 - \frac{R_\mathrm{cc}}{R_\mathrm{li}}\right).
\end{equation}

\noindent Considering $1 - R_\mathrm{cc}/R_\mathrm{li} \equiv \beta$ as a correction factor and using Eqs.~(\ref{Pouillet}) and (\ref{Liner_TCR_Eq_1}), the resistivity of the center conductor can be written as

\begin{equation}
\rho_\mathrm{cc} = \frac{\alpha}{\beta} R \left.\frac{d \rho_\mathrm{cc}}{d T}\right|_\mathrm{bulk} \left(\frac{dR}{dT}\right)^{-1}. \label{Liner_TCR_Eq_2}
\end{equation}

\begin{figure}[t]
  \centering
  \includegraphics[width=80 mm]{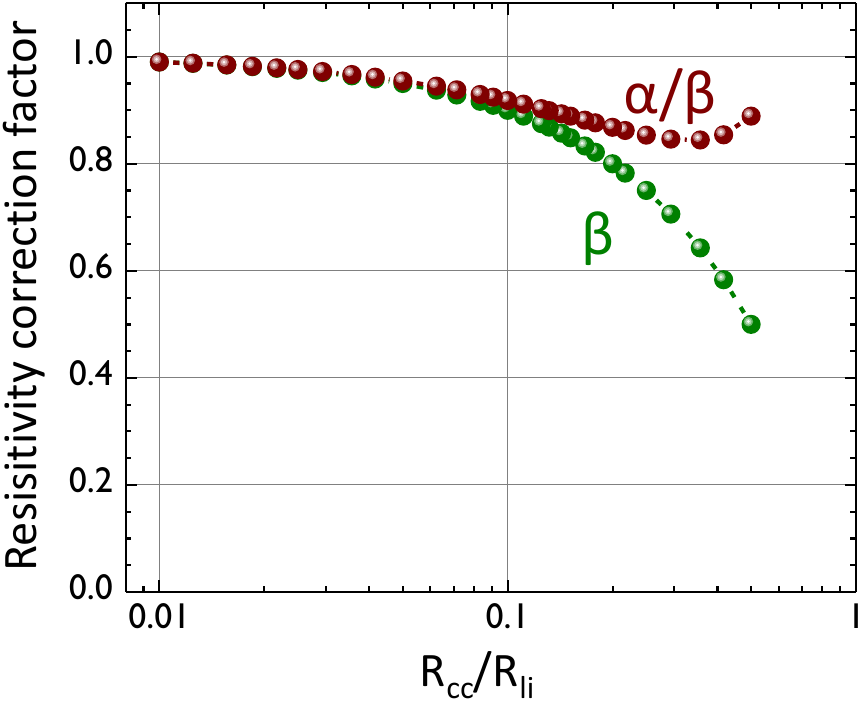}
\caption{\label{Fig_beta} Resistivity correction factor $\beta$ (green dots) as well as $\alpha / \beta$ assuming $d\rho_\mathrm{li}/dT = 0$ (brown dots).}
\end{figure}

\noindent Since $\beta \le 1$, the contribution of the barrier/liner conductance to the total wire conductance compensates partially for the overestimation of the electrical area. Figure~\ref{Fig_beta} shows values of $\beta$ as a function of $R_\mathrm{cc}/R_\mathrm{li}$. The figure also shows $\alpha /\beta$ as a function of $R_\mathrm{cc}/R_\mathrm{li}$ under the assumption that $d\rho_\mathrm{li}/dT = 0$. The results show that the resistivity $\propto \alpha /\beta$ is less affected than the electrical area. For all values of $R_\mathrm{cc}/R_\mathrm{li}$, the assumption of $\alpha /\beta = 1$, \textit{i.e.} neglecting the effect of the liner, leads to errors in the determination of $\rho_\mathrm{cc}$ of $< 20$\% at most and typically of less than 10\%. 

\section{Conclusions}

In conclusion, the applicability of the TCR method to scaled interconnect lines has been discussed addressing several aspects. For scaled interconnect wires with resistivities determined by surface and grain boundary scattering, the applicability of the central approximation of the TCR method, namely the substitution of the TCR of the interconnect wire by the bulk TCR becomes doubtful. Semiclassical calculations for thin films show that the TCR deviates from bulk values when the surface roughness scattering contribution to the total resistivity becomes non-negligible with respect to contributions of grain boundary scattering. For metallic thin films in the relevant thickness range with grain sizes comparable to the film thickness, the semiclassical model finds that grain boundary scattering typically dominates and the TCR is bulk-like, except maybe for ultrathin Cu films with large grains \cite{DSM_2017}. Alternative metals with shorter mean free paths than Cu should in general be less sensitive to surface scattering. Yet, surface roughness scattering can be expected to be enhanced in nanowires due to the larger surface-to-volume ratio and deviations from bulk TCR might be more important. The current lack of predictive models for the temperature dependence of nanowire resistivities including both surface roughness and grain boundary scattering prohibits a definitive conclusion and further work will be required to fully elucidate this issue. The use of measured thin film TCR values (instead of bulk TCR values) to determine cross-sectional area and resistivity of nanowires of the same material might improve the situation. However, in absence of predictive models for the nanowire TCR, the accuracy of this approach cannot currently be quantified. At a minimum, strongly non-bulk-like thin film TCR values indicate that the naive approximation of the nanowire TCR by the bulk value may be questionable and the results should be taken with caution and compared to physical cross-sectional areas.

In addition, to assess whether it can also be applied to rough interconnect lines, the TCR method has been rederived with an area that is allowed to vary along the line and an area-dependent resistivity that therefore also varies along the line. In such a case, it is shown that the TCR method yields the harmonic average of the area distribution function as well as, to first order, the correct value for the resistivity at the extracted area. A case study for Gaussian roughness indicates that the harmonic average is within a few \% of the arithmetic average for 3$\sigma$ LWR values that are comparable to what is achieved by current patterning techniques. Thus, the results of the TCR method are not expected to be affected for scaled interconnect lines with good yield.

Finally, the effect of a conductive barrier/liner layer on the TCR method has been discussed. The cross-sectional areas and resistivities of barrier/liner and center conductor cannot be determined independently by the TCR method. While the effect of the barrier/liner parallel conductance can be corrected for if it is accurately known, this will be the case only in rare situations. In current technology nodes, the contribution of the barrier and liners to the total conductance is still rather small and it can thus be treated as ``perturbation'' of the conductance of the center conductor. For such a case, the conditions, under which the assumption of a negligible barrier/liner contribution (``insulating'' barrier/liner) is valid, have been derived. Within this model, approximate correction factors can be calculated when approximate values for resistance, resistivity, and area ratios of center conductor and barrier/liner are known. The results show that neglecting barrier/liner resistance leads to an overestimation of the center conductor area that is approximately twice the relative conductivity contribution. An accuracy in cross-sectional area of 10\%{}  thus requires that the barrier/liner contributes less than 5\%{} to the total interconnect conductance. By contrast, the resistivity is less affected and remains typically within 10\%{} of the center conductor value even for relatively large relative barrier/liner conductance.

\section*{Acknowledgement}

The author would like to thank Shibesh Dutta (imec), Florin Ciubotaru (imec), Kristof Moors (University of Luxembourg) and Christian Witt (GlobalFoundries) for many valuable discussions. This work has been supported by imec's industrial affiliate program on nanointerconnects.

\end{document}